# WSe$_2$ Monolayers Grown by Molecular Beam Epitaxy on hBN

Julia Kucharek[1], Mateusz Raczyński[1], Rafał Bożek[1], Anna Kaleta[2], Bogusława Kurowska[2], Marta Bilska[2], Sławomir Kret[2], Takashi Taniguchi[3], Kenji Watanabe[4], Piotr Kossacki[1], Mateusz Goryca[1], Wojciech Pacuski[1]

[1] Institute of Experimental Physics, Faculty of Physics, University of Warsaw, Pasteura 5, 02-093 Warsaw, Poland,

[2] Institute of Physics, Polish Academy of Sciences, Aleja Lotników 32/46, 02-668 Warsaw, Poland,

[3] Research Center for Materials Nanoarchitectonics, National Institute for Materials Science, 1-1 Namiki, Tsukuba 305-0044, Japan

[4] Research Center for Electronic and Optical Materials, National Institute for Materials Science, 1-1 Namiki, Tsukuba 305-0044, Japan

Abstract: A three-step process was developed for growing high-quality, optically uniform WSe$_2$ monolayers by molecular beam epitaxy (MBE) with advantage of using hexagonal boron nitride (hBN). The process was optimized to maximize the efficiency of photoluminescence and promote formation of hexagonal WSe$_2$ domains. Atomic force microscopy (AFM) was employed to estimate the dispersion of WSe$_2$ hexagonal domains orientation. Monolayer character of the film was identified using optical methods and verified with high-resolution transmission electron microscopy (TEM) cross-section. Temperature-and-magnetic-field-dependent studies revealed the behaviour of exciton complexes to be analogical to that of exfoliated counterparts. Direct growth on hBN combined with uniform optical response proves this WSe$_2$ superior to mechanically exfoliated WSe$_2$ in terms of convenience of use and reproducibility. Provided results establish a significant progress in optical quality of epitaxially grown transition metal dichalcogenides (TMDs) monolayers and fabrication of large-scale functional devices.

Key words: Tungsten diselenide, molecular beam epitaxy, photoluminescence, atomic force microscopy, hexagonal boron nitride

## Introduction

Since the beginning of the graphene era, two-dimensional materials gained vast interest of electronics [1][2] and optoelectronics [3]. In particular, monolayers of Transition Metal Dichalcogenides from the MX$_2$ group attract attention due to their direct-gap, possibility of external dielectric screening tuning [4] and very strong Coulomb interactions [5]. WSe$_2$, owing its positive conduction band spin-splitting allows the observation of multi-particle excitonic states, particularly dark/grey states [6].

Although many successful bottom-up fabrications of WSe$_2$ materials have been realized (Table 1) the optical properties of WSe$_2$ are mostly studied on mechanically exfoliated layers encapsulated in hexagonal boron nitride (hBN) [7], owing to the efficiency of their optical response. So far, mechanical transfer and encapsulation in hBN was needed also for improvement of optical quality of CVD grown WSe$_2$ monolayer [8].

| Technique/ | MBE | CVD |
|---|---|---|
| /Substrate | Sapphire [9]<br>Mica [11]<br>AlN [13]<br>HOPG [16, 17]<br>Bilayer graphene [18]<br>Epitaxial graphene on SiC [20]<br>ZrO [22]<br>Au [24]<br>GaP [26]<br>SrTiO3 [27] | Sapphire [10]<br>YIG [12]<br>SiO2 [14, 15]<br>Si3N4 [15]<br>Epitaxial graphene [19]<br>Fused silica [21]<br>Au [23]<br>hBN [25] |

*Table 1. Overview of $WSe_2$ growth on different substrates by MBE and CVD techniques.*

In practice the most desired materials are these that can be produced in a repeatable, schematized manner, resulting in large uniform layers with predictable properties. While exfoliated materials can be analysed to find a particular working spot that has desired optical properties, the same cannot be reliably said about its neighbourhood.

Addressing these demands, we present tungsten diselenide grown by molecular beam epitaxy on hexagonal boron nitride exhibiting high-quality optical properties, uniform over dozens of squared micrometres. Our $WSe_2$ exhibits photoluminescence (PL) signal with a set of emission lines originating from multi-particle excitonic states, with no need of upper hBN coverage. MBE technique enables a well-defined, reproducible growth process directly on hBN substrate. HBN flakes as a substrate for TMD growth present many advantages e.g. smooth surface without dangling bonds and uncompensated charges facilitates atoms surface mobility. Atomic-terrace-free surface promotes the continuous growth of the TMDs film. Chemical and high-temperature resistance make hBN a convenient substrate which doesn't require complicated preparations before growth. 90 nm of $SiO_2$ and 10-20 nm of hBN [28] effectively decreases inhomogeneous broadening of the spectral lines [29] improving optical quality of the material.

In previous works, we have shown a successful results on growth of bright $MoSe_2$ monolayers on hBN [30, 31]. Notably since $WSe_2$ is a dark material, very sensitive to deficiency of selenium, it requires precise adjustments of growth parameters to achieve comparable results.

**Samples**

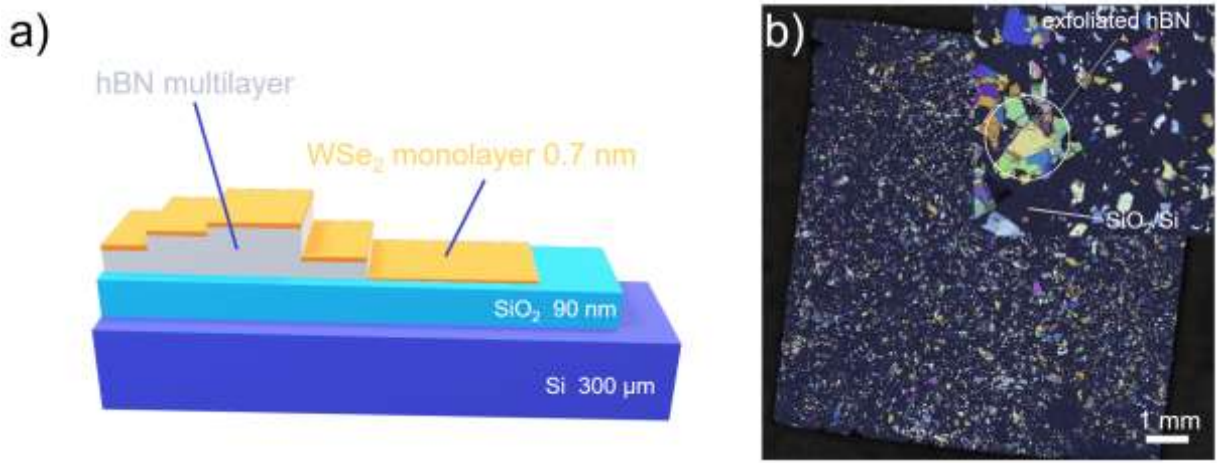

*Figure 1. a) Scheme of a sample. b) Optical microscope image of the sample.*

In this article, we introduce molecular beam epitaxy growth on hBN for high-quality, optically uniform, large-scale WSe$_2$ monolayers. The growth was realized in three-step process including : pre-growth annealing to outgas the surface and flatten hBN flakes, slow growth phase and post-growth annealing facilitating tungsten atoms rearrangement and hence improving the crystalline order of the layer.

In order to grow WSe$_2$, intrinsic (001) Si wafer with SiO$_2$ of 90 nm was prepared in a chemical cleaning process, and then exfoliated hexagonal boron nitride was transferred onto it forming a complete substrate for the growth. The scheme and the optical image of the sample are depicted in Figure 1.

Before growth, the substrate was annealed in 700°C for 1h to outgas the surface and relax strain in exfoliated hBN flakes, parallelly Se source was warming up to reach the growth target flux with Se shutter open. Growth was performed at 300°C for 2.5 h and 3 h respectively for the two samples (UW2144 and UW2145) used for this paper. Tungsten e-beam source power set at 150 W corresponded to 0.3 Å/h tungsten deposition rate. During growth selenium deposition rate is 8 × 10$^3$ Å/h and changes to 4 × 10$^4$ Å/h for the last step of the process. After-growth-annealing was performed at 800°C for 2 h. Finished WSe$_2$ samples were removed from the growth chamber right after reaching the growth temperature for the second time, in order to avoid selenium condensation on the surface.

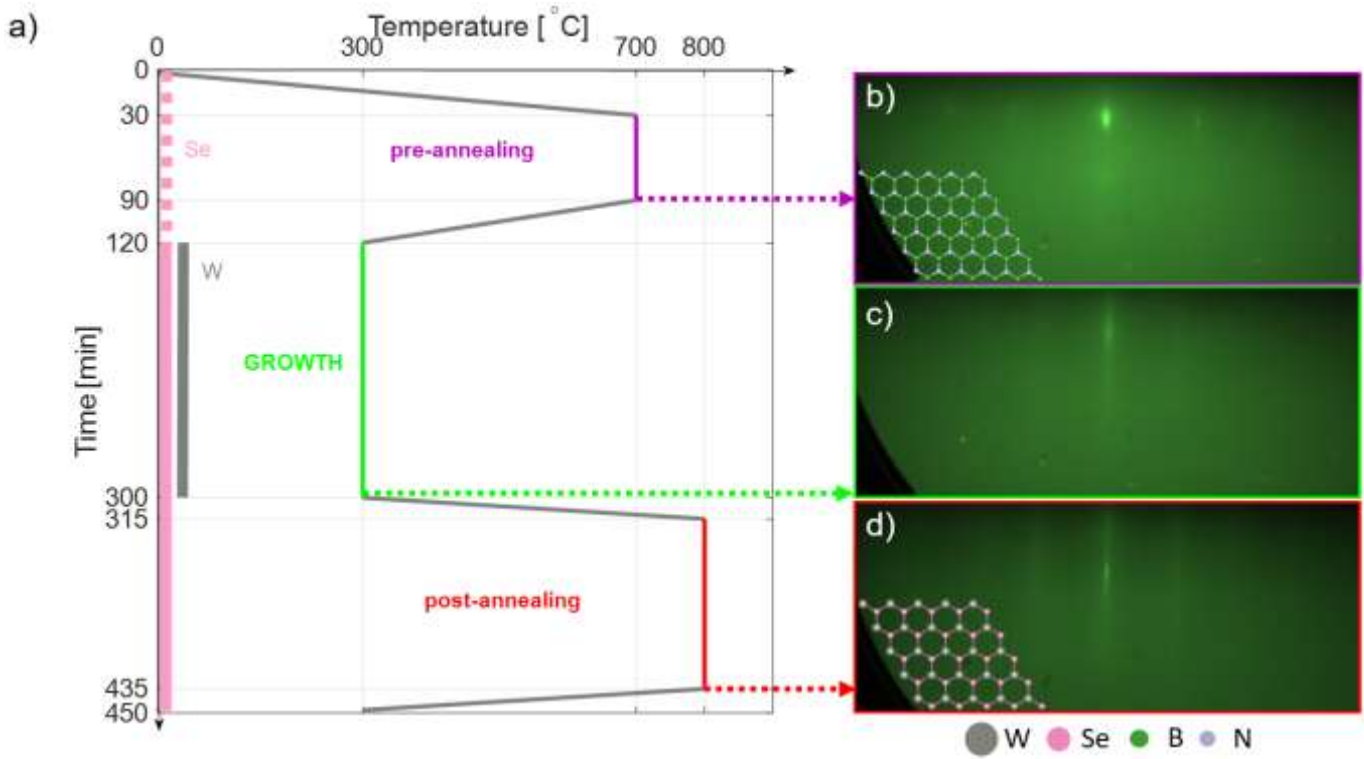

Figure 2. a) Scheme of the growth process presenting substrate temperature evolution in time. b)-d) RHEED pattern of the annealed substrate before growth, WSe$_2$ sample after growth, and after final annealing respectively.

Reflection High Energy Electron Diffraction (RHEED) was used to monitor the growth in situ. In the Figure 2 b)-d) every step of the growth is documented with RHEED images. After pre-growth annealing hBN lines appear, following the slow growth phase, only WSe$_2$ lines are visible, and after post-growth annealing the lines are sharper and brighter which suggests higher crystalline order. WSe$_2$ strikes align with the ones of hBN indicating epitaxial orientation of TMD domains with regards to the substrate. Assuming the initial hBN lattice constant to be 2.50 Å [32], one can estimate the WSe$_2$ lattice constant to be 3.24 Å, which is less than 3.282 Å measured in [33], but close to relaxed monolayer a=b lattice constant equal 3.247 Å claimed in [34] theoretical calculations.

**Results**

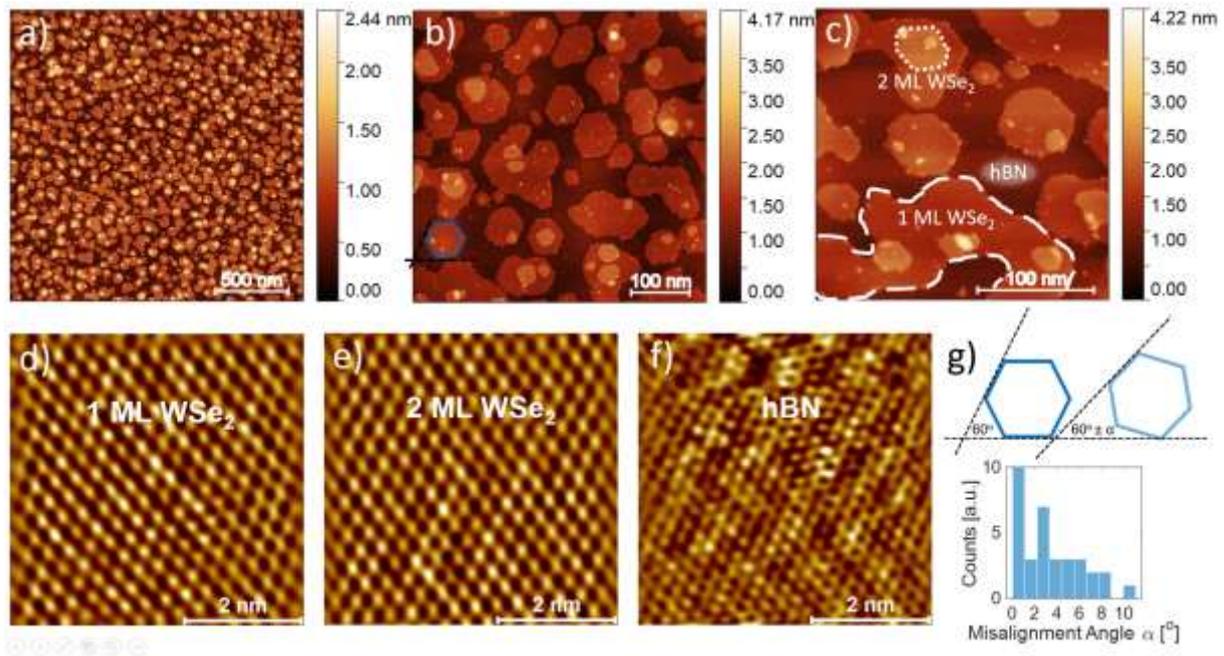

*Figure 3. a)-c) Atomic Force Microscopy images of UW2144 sample in three different magnifications. In c). monolayer, bilayer, and uncovered hBN areas are marked. d)-f) High-Resolution Atomic Force Microscopy (HRAFM) images of UW2144 sample surface. g) – scheme showing the flakes' misalignment angle α and histogram basing on b).*

For structure investigations of $WSe_2$ single crystal film, we have used ambient AFM. Atomic resolution images were acquired in Ultra High Vacuum using Contact Mode AFM (UHVAFM). Figure 3 shows partially coalesced $WSe_2$ flakes grown by MBE on hBN substrate. The upper row displays pictures in 3 magnifications. The first one (a) shows the general coverage of the surface of hBN to be 55 %. Moreover, only 16 % of the monolayer is covered with a second or more layers. The average roughness of $WSe_2$ surface in a) is 184 pm.

The tendency of $WSe_2$ to form hexagonal flakes is well demonstrated in b), proving that both W and Se edges in our material are equal and well formed. In places where $WSe_2$ forms bigger, coalesced flakes, one can also find 120° angles between edges. For each flake with at least one defined corner and straight edge, the average misalignment angle α of the flakes is 3.4°, marked in Fig. 3g).

$WSe_2$ nucleation centres appear on both the hBN (between the $WSe_2$ flakes) and on the already-grown $WSe_2$ flakes (Fig. 3c). There is no strong trend, nucleation centres are dispersed equally all over the area. Additionally, exemplary zones of monolayer and bilayer of $WSe_2$ and uncovered hBN are marked. The lower row shows UHVAFM images of the monolayer, a bilayer of $WSe_2$, and a multilayer of hBN subsequently. Lattice constants of the materials were calculated as an average of 10 units measured along the three apical symmetry axes going through hexagon corners. The estimation of $WSe_2$ and hBN lattice constant along the a=b axis is 3.51 Å and 2.71 Å for $WSe_2$ (both monolayer and bilayer) and hBN respectively.

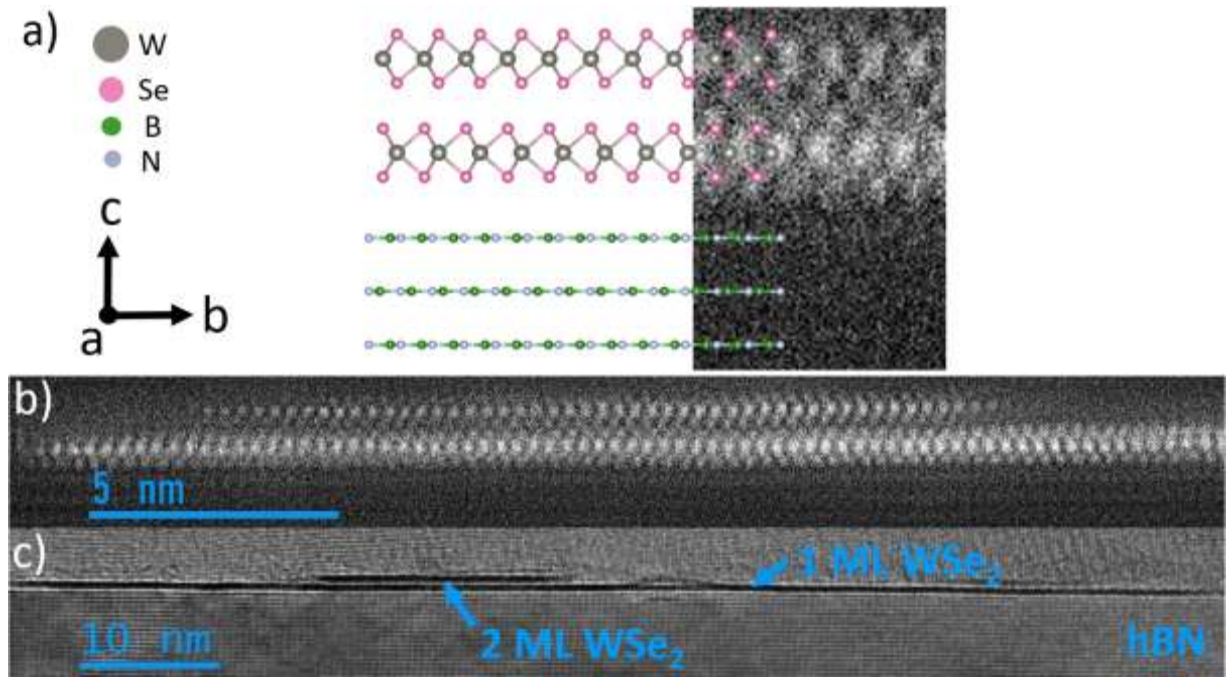

*Figure 4. Transmission Electron Microscopy images of WSe$_2$ cross-section. a) Model presenting WSe$_2$ atomic structure combined with a dark-field image. b) Dark-field image. c) Bright-field image.*

Further characterisation with TEM allows to assess the stacking regime of grown material. Cross-sections of WSe$_2$ monolayer and bilayer regions are shown. Panel a) presents the 2H structure of two subsequent WSe$_2$ layers. The atomic structure inset follows an atom arrangement. In panel c) a 100 nm long monolayer is grown on a well-ordered hBN crystal. There is also an area of about 20 nm, where the monolayer is covered with a second layer. Even though WSe$_2$ crystal doesn't form a complete monolayer, film is continuous over hundreds of nanometres. The bright – field image c) clearly reveals exfoliated hBN structure. TEM images confirm van der Waals growth of both, first of the WSe$_2$ layer on hBN followed by WSe$_2$ bilayers.

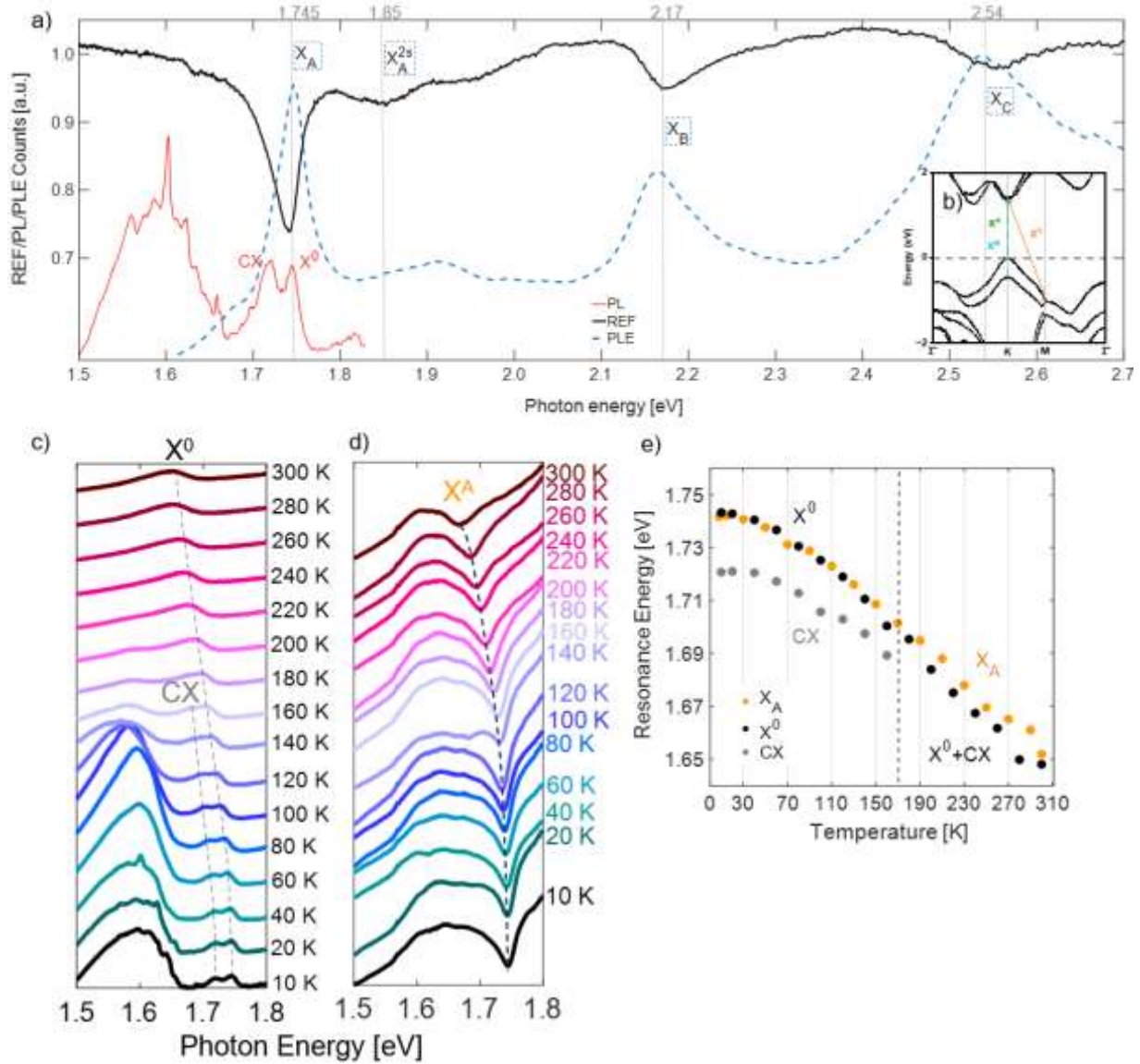

Figure 5. a) Exemplary photoluminescence, reflectance, and photoluminescence excitation spectra of $WSe_2$. b) Exciton A, B and C transitions in $WSe_2$ band structure [35, 36]. c) Neutral and charged exciton PL energies temperature dependency. d) Exciton A absorption temperature dependency. e) Comparison of exciton A absorption energy dependence with neutral and charged excitons PL energy. 532 nm laser with approximately 200 µW / µm² power density was used.

To show MBE-grown $WSe_2$ optical response characteristics µ-photoluminescence, µ-photoluminescence excitation (PLE) and µ-reflectance (REF) were studied. In Figure 5 a) PL spectrum (red line) shows a wide range of sharp lines characteristic of $WSe_2$ monolayer. Neutral (X0) and charged excitons (CX) on 1.745 eV and 1.72 eV respectively, are marked. The black line, referring to the reflectance spectrum shows exciton A absorption on 1.75 eV, and excitons B and C located at 2.17 eV and 2.54 eV. The absorption minimum on 1.85 eV corresponds with exciton A 2s state. Dashed (blue line) shows the PLE measurement, confirming the appearance of the positions of A, $A^{2s}$, B, and C excitons. Zero-stokes shift between PL and REF is confirmed by low-temperature (Fig. 5a) and temperature-dependent study (Fig.5e).

For both PL and REF, we observe the temperature-dependent redshift and broadening of the lines. Analogically to exfoliated counterpart, MBE-$WSe_2$ reveals more intense emission from localized/charged exciton states in lower temperatures (10 – 140 K) and this trend reverses for temperatures higher than 160 K. This is in contrast to exfoliated materials where the integrated PL intensity is increasing monotonously with temperature.

X0 energy (or exciton A) changes in temperature from 1.745 eV (in 10 K) to 1.649 eV (in 300 K) and reveals overlapping results with those published in A. Aurora et al. [35]. Comparison of both is shown in fig. S3.

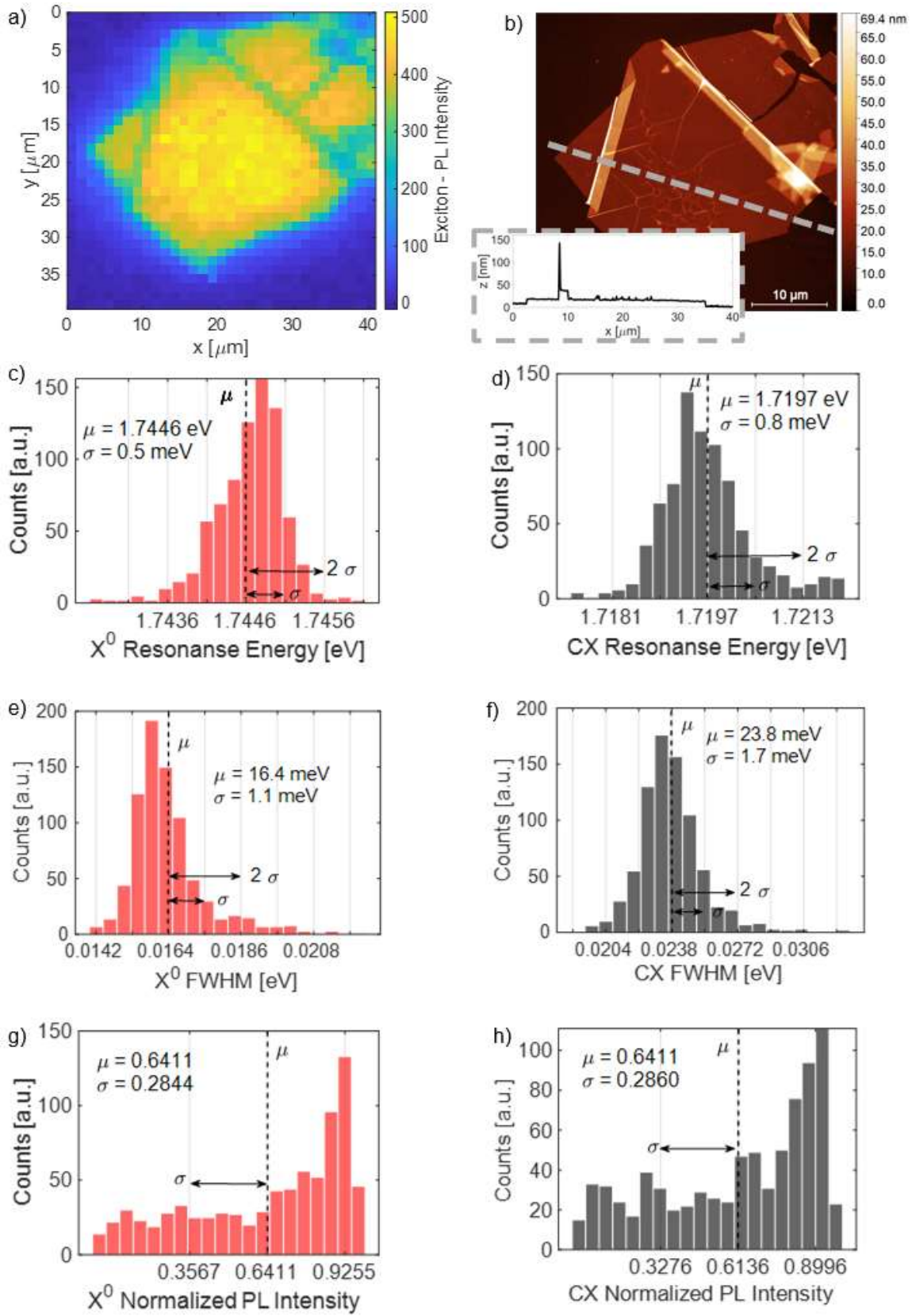

*Figure 6. WSe$_2$ grown on hBN low-temperature (10K) µ-PL map and statistics. a) Neutral exciton PL intensity map. b) hBN flake mapped in a). Histograms of neutral and charged excitons fitted parameters. Left column is dedicated to neutral exciton and right one to charged exciton. c) and d) – resonance energy positions. e) and f) – full widths at halt maximums. g) and f) – normalized PL intensities. On each histogram average value and standard deviation are marked.*

To reveal the exceptional optical homogeneity of the material, we show spatially-resolved PL measurements (Figure 6 a)), where the intensity of the PL signal of X0 is displayed. The PL map precisely mirrors the measured hBN flake presented in Figure 6 b). It is worth mentioning that in Figure 6 a) PL intensity changes with respect to hBN thickness, and to clarify this effect the small inset in c) with cross-section of studied flake is placed. The PL signal is similar uniformly on the surface of the ~10 nm thick hBN flake. The map is followed by statistical analysis of the PL spectra collected on the hBN flake. All the statistics exclude spectra from the outside of the flake, but take into account all of the spots where hBN is inhomogeneous (e.g. wrinkles, edges, cracks or folds). By fitting Gaussian function (due to significant inhomogeneous broadening), we received neutral and charged excitons resonance energies, full widths at half maximum (FWHM) and normalized excitons energies.

Average positions of X0 $\mu_{X0}$ = 1.7446 eV and CX $\mu_{CX}$ = 1.7197 eV are in agreement with values published for exfoliated materials encapsulated in hBN measured in low temperature PL [37–39]. Standard deviation and FWHM of the charged exciton reach higher values than those for neutral exciton, 0.5 meV and 0.8 meV, 16.4 meV and 23.8 meV respectively. We can attribute it to the fact that X0 resonance is composed of one emission line and CX consists of two emission lines: spin-singlet and spin-triplet trion [40]. However, average linewidth for X0 and CX is notable, we find lines as narrow as 10 meV for X0 and 15 meV for CX, which is 1.5 - 3 times more than for near-homogeneous exfoliated (hBN-encapsulated) WSe$_2$ linewidths [41]. Using hBN with more invariable surface morphology could reduce the inhomogeneous broadening to some extent.

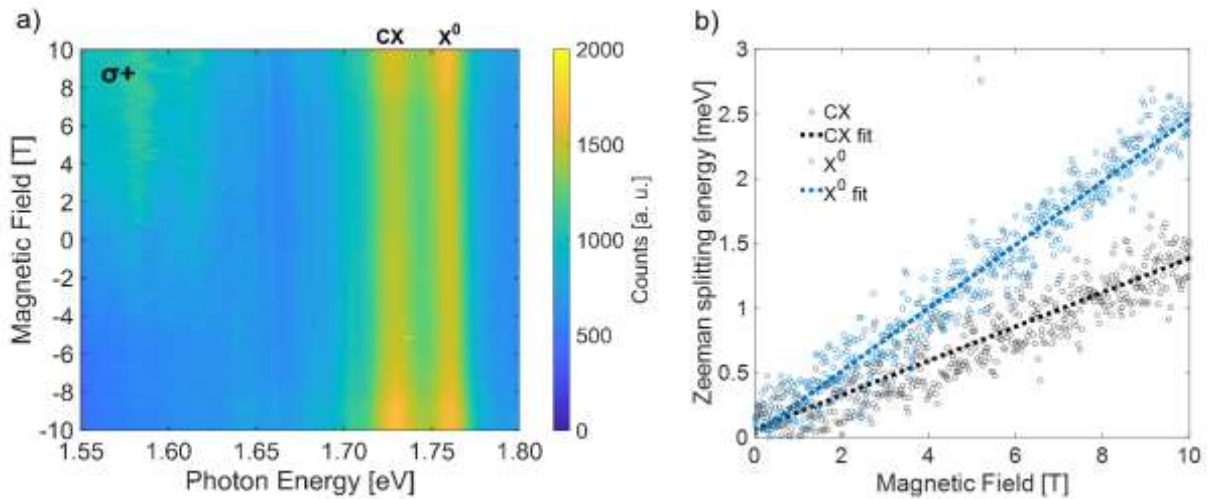

*Figure 7. a) σ+ polarization in detection of PL in function of magnetic field. b) Zeeman splitting energy of neutral and charged exciton with g-factors of -4.4 for X0 and -2.4 for CX. Measurements were performed in low-temperature (10 K) using a 635 nm laser for excitation.*

Figure 7. a) displays the behaviour of PL response under an applied magnetic field. In the picture σ+ polarization is shown. Besides two main lines of X0 and CX, a branch of states in lower energies is also present, all redshifting in increasing magnetic field. The Zeeman splitting energy of X0 and CX is also demonstrated in Figure 7. b). Effective Lande g-factors for X0 and CX are - 4.4 and - 2.4 respectively. When the g-factor of the neutral exciton aligns with expected values, the measured trion g-factor is notably lower than typical values reported for exfoliated WSe$_2$. Due to the inability to spectrally resolve singlet (SST) and triplet (TST) spin states of trions, the magnetic-field dependence of oscillator strength becomes a critical factor. Assuming analogous behavior to that observed in WS$_2$ (as reported in [42]), the oscillator

strength of the SST increases under σ+ polarization, while that of the TST decreases at a corresponding rate. This trend is reversed for σ− polarization.

If the variation in oscillator strength is taken into account, and assuming a highly symmetric configuration, the effective trion g-factor is expected to be approximately halfway, yielding a value around –2.2. This estimation is in good agreement with our experimental observations.

**Discussion**

The literature reports shows that among epitaxial methods, CVD is mostly exploited for TMDs growth. To compare with MBE grown $WSe_2$ on hBN, the results from Zhang X. et al. [25] were chosen as representative of CVD, due to their leading film quality. Comparing exciton and trion peaks in 80 K with our results, one can see that X0 FWHMs are similar: 18.6 meV and 19.2 meV. CX FWHMs – 38.5 meV and 19.8 meV in our case. Authors of [25] claim electron binding energy in trion to be ~30 meV, we can estimate our close to 25 meV. Smaller binding energy can be attributed to low intrinsic doping or smaller $WSe_2$ flakes which can lead to shorter lifetime of excitons. Longer edge line can also explain appearance of multiple lines originating from localised states in the spectrum between 1.50 and 1.65 eV which stays in contrast to that of [25]. Notably, in both publications, the decrease of integral PL intensity with temperature is observable between 80 and 300 K. Being a dark material, $WSe_2$ should present the opposite tendency, as shown for mechanically exfoliated monolayers [35]. This unexpected behaviour suggests changes in radiative/nonradiative channelling mechanisms possibly due to strain from growth on hBN substrate, and states attractive platform for further investigations.

Despite the fact that an extensive progress in MBE-fabrication of $WSe_2$ was introduced, there is plenty of room for further improvements. Among the necessary changes, we would like to place the optimization of the process for continuous layer growth suppressing multilayer formation. $WSe_2$ sample presented in this work has a hBN surface coverage of 55%. Above this value, we observe significant multilayer growth and after exceeding a certain amount of material, excitonic lines are not resolved anymore and create one broad band (Fig.S1.). Therefore, actions towards structurally uniform $WSe_2$ layer are inevitable to expand applications of our material to transport studies or use in large-area heterostructures.

In exfoliated materials, encapsulating is indispensable to the quality of optical measurements, our experiments show that covering as-grown $WSe_2$ with hBN is not necessary and doesn't further improve the results. The intensity of PL and the linewidths stay the same, while X0 and CX lines shift slightly towards lower energies (Fig.S2).

Moreover, exfoliated hBN flakes are inherently size-limited; in order to achieve a functional size of the final product, we need to find a source of large uniform hBN layers. This is already partially achieved with MOCVD grown hBN [43], however while it can reach inches of area, the uniformity of the surface is lacking, compromising the integrity of TMDs layers grown atop. Further improvements to this material are key in developing next-gen opto-electronic devices.

**Conclusions**

We have showcased a three-step growth process of $WSe_2$ by molecular beam epitaxy on exfoliated hBN flakes, which gives a tool for reproducible production of optically uniform monolayers. Through carefully chosen temperatures, durations and W:Se flux ratios for each of the three steps, we have, for the first time, achieved high optical effectiveness in $WSe_2$ epitaxial films. We have accomplished optical response, comparable to exfoliated counterparts without a need to fully encapsulate WSe$_2$ with upper hBN. Crucially, we have developed WSe$_2$ optical uniformity within hundreds of µm$^2$ exhibiting high durability without capping. The samples preserve optical properties for over one year if kept in moderate negative pressure - 0.1 Pa. Due to these improvements, for the first time MBE-grown WSe$_2$ has displayed clearly resolved neutral and charged excitons lines as well as localised and charged states. Behaviour of those states was observed in a function of external magnetic field and temperature. AFM images show hexagonal structure of the layer to be free from major defects. By TEM we have proved growth in 2H van der Waals regime. The integrity of the results present a highly valuable advancement in bottom-up production of 2D TMDs.

## Methods

### Substrate preparation

The Si substrate (intrinsic semiconductor) with SiO$_2$ of 90 nm was cleaved into smaller pieces and washed using ultrasonic cleaner in a three-step process. Firstly it was sonicated in isopropanol, then acetone, and the end in demineralized water. After that, it was backed at 200 degrees on the hotplate for 10 min. Directly after that, the substrates were stuck to the tape with exfoliated hBN flakes and covered with a spare piece of foil to protect them from dust. Right before the growth substrates were detached from the foil and placed in the MBE.

### Molecular Beam Epitaxy

Molecular Beam Epitaxy reactor by SVT Associates, Inc. The growth was conducted in chamber dedicated to II-VI materials, under an ultra-high vacuum with a background pressure below $1 \times 10^{-9}$ Torr. Tungsten e-beam source and selenium Knudsen cell were employed with source elements of tungsten in rod form (99.98% purity, 2 mm diameter, 35 mm length) and selenium granules (99.99999% purity). The MBE Reactor was accompanied by Reflection High Electron Energy Diffraction (RHEED). The electron accelerating potential used was 8 kV.

### (Ultra High Vacuum UHV) Atomic Force Microscopy AFM

High Resolution Atomic Force Microscopy measurements were performed in vacuum chamber (~$5 \times 10^{-10}$ mbar) with an Omicron VT XA microscope in contact mode using NT-MDT CSG10 cantilevers. Results were processed using Scanning Probe Image Processor v.6.7.9 (Image Metrology A/S) software.

AFM images were obtained using Bruker Icon Dimensions microscope with Nanoscope VI controller. Peak Force Tapping mode with ScanAsyst-Air probe was employed.

### Transmission Electron Microscopy

The size of the largest measured hBN flakes was around 40-60 microns. Cross-sections were made using a focused gallium ion beam with a HELIOS Nanolab 600 dual-beam electron-ion microscope. Lamellas were prepared both directions: parallel to the flake edges and parallel to the cracks visible in SEM images.

Before the FIB process, a ~10 nm thick carbon layer was deposited on the flake surface to prevent charging during SEM imaging and the deposition of a protective platinum layer via electron-beam deposition from GIS platinum source. The standard method of cutting and lift-out of the lamella using the Omniprobe nanomanipulator was applied.

During the thinning process to achieve electron transparency, an unusual problem occurred: the protective platinum curled, masking the thin part of the lamella. This issue is associated with the presence of metallic selenium above the $WSe_2$ layer, which has poor mechanical strength compared to hBN and platinum, and likely exhibits chemical affinity for gallium.

As a result, the standard final thinning procedure was modified. Only small areas were thinned to a thickness suitable for high-resolution imaging (max up to 100 nm long), which prevented the protective platinum from curling but significantly reduced the useful observation area to just a few nanometers.

FIB cross-sectional imaging of the $WSe_2$ layers was performed in AC-HRTEM (Aberration-Corrected High-Resolution Transmission Electron Microscopy) mode and HR-STEM mode using a HAADF detector, with a 76 mm camera length providing Z-contrast proportional to the average atomic number of the atomic columns. All TEM studies were conducted using a FEI Titan Cube 80-300 microscope operating at 300 kV.

*Optical study*

Photoluminescence (PL), Reflection (Refl), and Photoluminescence Excitation (PLE) measurements were performed in a flow helium cryostat (Lakeshore Janis ST–500) at the temperature of 10 K. To reach a micrometer spatial resolution better than the exfoliated h--BN flake sizes the Nikon L Plan objective was used. Its magnification was x100 with NA = 0.7 and the working distance of 6.5 mm. The light source used during the reflectivity measurement was a halogen lamp filtered spatially via a pinhole and set of lenses. The PL excitation source was a CW 532 nm semiconductor laser (Omicron LightHUB COBOLT Samba 100 mW) and the power at the sample position was kept close to 200 µW. Finally, for performing a PLE measurement a white light supercontinuum laser (the SuperK EXTREME) together with a monochromator (the LLTF Contrast VIS HP8) providing an output light spectrum narrower than 2.5nm (FWHM). To detect light coming back from the sample the spectrometer Andor SR-500i was used with the diffraction grating 600/500nm and Peltier cooled CCD camera of Andor model DV420-FI.

The determination of excitonic Lande g-factors was performed in a helium-bath cryostat, in a gaseous helium atmosphere at 10 K in PL configuration. The cryostat was Oxford Spectromag 4000 with a superconducting magnet reaching a field up to 10 T. The excitation was CW 532 mn semiconductor laser with well-defined linear polarisation. The simultaneous detection in both σ+ and σ- polarisation was achieved by using quarter-waveplate and Wollaston prism deflecting the light along the spectrometer entrance slit. Spectrometer was SpectraPro HRS-300 (Princeton Instruments) with grating 600/500nm and the CCD matrix camera iDus DU420A-OE (Andor).

**Acknowledgements**

We acknowledge the financial support from National Science Centre Poland project no. 2021/41/B/ST3/04183 and 2023/49/N/ST11/04125. K.W. and T.T. acknowledge support from the JSPS KAKENHI (grant numbers 21H05233 and 23H02052) and the World Premier International Research Center Initiative (WPI), MEXT, Japan.## References

[1] C.-H. Lee, G.-H. Lee, A.M. van der Zande, W. Chen, Y. Li, M. Han, X. Cui, G. Arefe, C. Nuckolls, T.F. Heinz, J. Guo, J. Hone, P. Kim, *Nature nanotechnology* **9**, 676 (2014).

[2] A. Nourbakhsh, A. Zubair, R.N. Sajjad, A. Tavakkoli K G, W. Chen, S. Fang, X. Ling, J. Kong, M.S. Dresselhaus, E. Kaxiras, K.K. Berggren, D. Antoniadis, T. Palacios, *Nano letters* **16**, 7798 (2016).

[3] T. Tan, X. Jiang, C. Wang, B. Yao, H. Zhang, *Advanced science (Weinheim, Baden-Wurttemberg, Germany)* **7**, 2000058 (2020).

[4] D. Tebbe, M. Schütte, K. Watanabe, T. Taniguchi, C. Stampfer, B. Beschoten, L. Waldecker, *npj 2D Mater Appl* **7** (2023).

[5] S. Brem, M. Selig, G. Berghaeuser, E. Malic, *Scientific reports* **8**, 8238 (2018).

[6] X.-X. Zhang, Y. You, S.Y.F. Zhao, T.F. Heinz, *Physical review letters* **115**, 257403 (2015).

[7] F. Cadiz, E. Courtade, C. Robert, G. Wang, Y. Shen, H. Cai, T. Taniguchi, K. Watanabe, H. Carrere, D. Lagarde, M. Manca, T. Amand, P. Renucci, S. Tongay, X. Marie, B. Urbaszek, *Phys. Rev. X* **7** (2017).

[8] A. Delhomme, G. Butseraen, B. Zheng, L. Marty, V. Bouchiat, M.R. Molas, A. Pan, K. Watanabe, T. Taniguchi, A. Ouerghi, J. Renard, C. Faugeras, *Appl. Phys. Lett.* **114** (2019).

[9] M. Nakano, Y. Wang, Y. Kashiwabara, H. Matsuoka, Y. Iwasa, *Nano letters* **17**, 5595 (2017).

[10] X. Zhang, T.H. Choudhury, M. Chubarov, Y. Xiang, B. Jariwala, F. Zhang, N. Alem, G.-C. Wang, J.A. Robinson, J.M. Redwing, *Nano letters* **18**, 1049 (2018).

[11] C. Vergnaud, M.-T. Dau, B. Grévin, C. Licitra, A. Marty, H. Okuno, M. Jamet, *Nanotechnology* **31**, 255602 (2020).

[12] Y. Ren, L. Yang, M.-C. Wang, Y. Lv, Y. Fan, M. Abid, C.Ó. Coileáin, C.-R. Chang, H.-C. Wu, *J. Phys. Chem. C* **128**, 5228 (2024).

[13] K.E. Aretouli, D. Tsoutsou, P. Tsipas, J. Marquez-Velasco, S. Aminalragia Giamini, N. Kelaidis, V. Psycharis, A. Dimoulas, *ACS applied materials & interfaces* **8**, 23222 (2016).

[14] B. Liu, M. Fathi, L. Chen, A. Abbas, Y. Ma, C. Zhou, *ACS nano* **9**, 6119 (2015).

[15] *Chemical Vapor Deposition of Large-Sized Hexagonal WSe2 Crystals on Dielectric Substrates* 2015.

[16] R. Yue, Y. Nie, L.A. Walsh, R. Addou, C. Liang, N. Lu, A.T. Barton, H. Zhu, Z. Che, D. Barrera, L. Cheng, P.-R. Cha, Y.J. Chabal, J.W.P. Hsu, J. Kim, M.J. Kim, L. Colombo, R.M. Wallace, K. Cho, C.L. Hinkle, *2D Mater.* **4**, 45019 (2017).

[17] J.H. Park, S. Vishwanath, X. Liu, H. Zhou, S.M. Eichfeld, S.K. Fullerton-Shirey, J.A. Robinson, R.M. Feenstra, J. Furdyna, D. Jena, H.G. Xing, A.C. Kummel, *ACS nano* **10**, 4258 (2016).

[18] Y. Zhang, M.M. Ugeda, C. Jin, S.-F. Shi, A.J. Bradley, A. Martín-Recio, H. Ryu, J. Kim, S. Tang, Y. Kim, B. Zhou, C. Hwang, Y. Chen, F. Wang, M.F. Crommie, Z. Hussain, Z.-X. Shen, S.-K. Mo, *Nano letters* **16**, 2485 (2016).

# SUPPLEMENBTARY INFORMATION

# WSe$_2$ Monolayers Grown by Molecular Beam Epitaxy on hBN with Exceptional Optical Uniformity

Julia Kucharek[1], Mateusz Raczyński[1], Rafał Bożek[1], Anna Kaleta[2], Bogusława Kurowska[2], Marta Bilska[2], Sławomir Kret[2], Takashi Taniguchi[3], Kenji Watanabe[4], Piotr Kossacki[1], Mateusz Goryca[1], Wojciech Pacuski[1]

[1] Institute of Experimental Physics, Faculty of Physics, University of Warsaw, Pasteura 5, 02-093 Warsaw, Poland,

[2] Institute of Physics, Polish Academy of Sciences, Aleja Lotników 32/46, 02-668 Warsaw, Poland,

[3] Research Center for Materials Nanoarchitectonics, National Institute for Materials Science, 1-1 Namiki, Tsukuba 305-0044, Japan

[4] Research Center for Electronic and Optical Materials, National Institute for Materials Science, 1-1 Namiki, Tsukuba 305-0044, Japan

| Sample | Growth temperature [°C] | Growth time [h] | Annealing temperature [°C] | Annealing time [h] |
|---|---|---|---|---|
| UW2144 | 300 | 2.5 | 800 | 2 |
| UW2145 | 300 | 3 | 800 | 2 |

*Table S1. Growth details of samples UW2144 and UW2145 investigated in the main text: growth temperature and time, annealing temperature and time.*

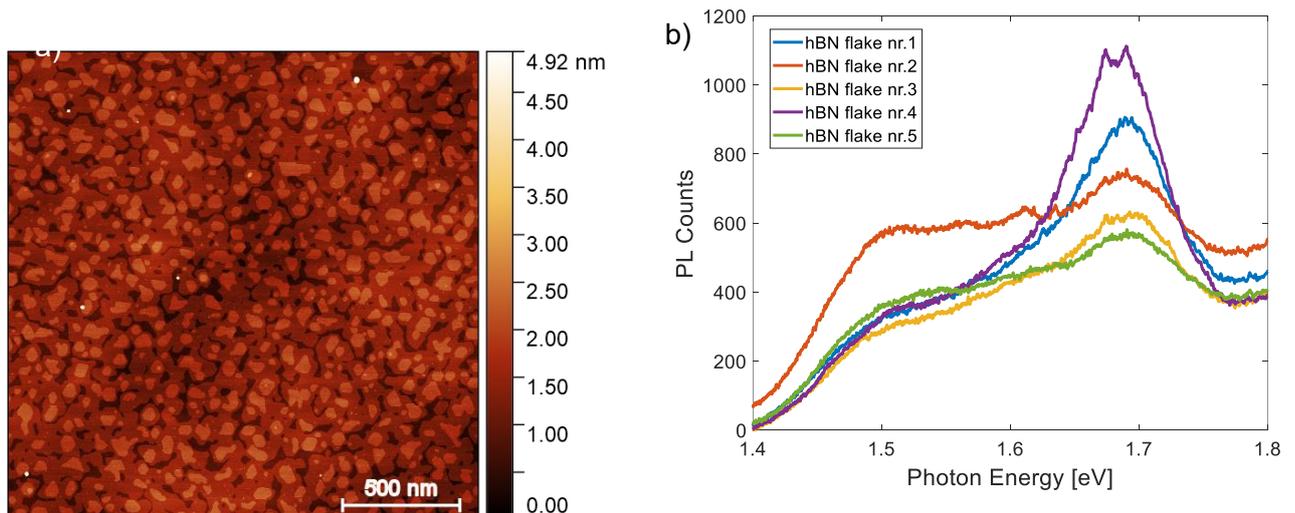

*Figure S1. a) AFM image of sample (UW2083) exceeding 70% of hBN surface coverage. Monolayer plane is nearly coalesced and significant part of the film is composed of bilayer. b) PL spectra for a) measured in four different places on the sample. All shows that PL signal in the region of neutral and charged excitons reveals one broad band with a maximum ~1.69 eV, below charged exciton energy. Observed emission energy is higher than the one from bilayer noticed in literature, but it is too low to be created directly from connection of exciton and trion lines. 515 nm laser with power of 800 µW was used. Measurement temperature was 10 K.*

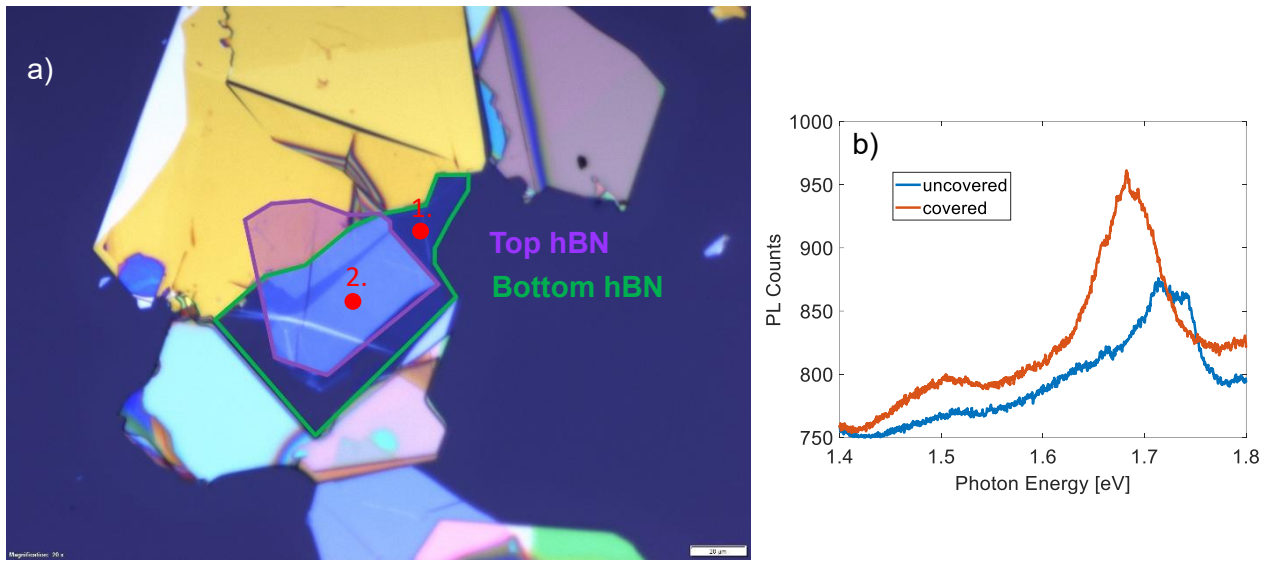

*Figure S2. a) Optical image of WSe$_2$ film grown by MBE on hBN (UW2266) covered with top hBN through polymer stamp dry method. After the transfer, the sample was re-annealed using the same procedure as for post-growth annealing. Spots 1. (uncovered) and 2. (covered) were measured for comparison after finishing the entire process. b) PL spectra of covered (orange) and uncovered (blue) regions. 532 nm laser with power of 600 µW was used. Measurement temperature was 10 K.*

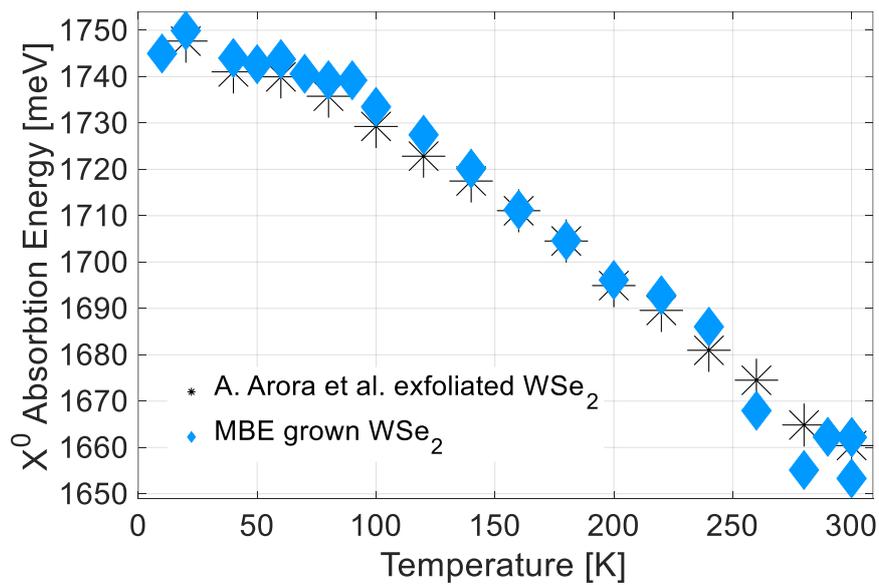

*Figure S3. Comparison of X$^0$ evolution in temperature. Black stars are points measured by A. Aurora et al. [1] on exfoliated WSe$_2$ encapsulated in hBN, blue diamonds are results from our MBE-grown WSe$_2$. Both in agreement with each other.*